\documentclass[10pt]{article}
\usepackage{graphicx}
\def\Title#1{\begin{center} {\Large #1 } \end{center}}
\def\Author#1{\begin{center}{ \sc #1} \end{center}}
\def\Address#1{\begin{center}{ \it #1} \end{center}}

\newcommand\pubblock{\rightline{\begin{tabular}{l} Proceedings of the CTD/WIT 2019\\ \pubnumber\\
         \pubdate  \end{tabular}}}

\newenvironment{Abstract}{\begin{quotation} \begin{center} 
             \large ABSTRACT \end{center}\bigskip 
      \begin{center}\begin{large}}{\end{large}\end{center} \end{quotation}}

\newenvironment{Presented}{\begin{quotation} \begin{center} 
             PRESENTED AT\end{center}\bigskip 
      \begin{center}\begin{large}}{\end{large}\end{center} \end{quotation}}

\def\beq{\begin{equation}}
\def\eeq#1{\label{#1}\end{equation}}
\def\eeqn{\end{equation}}

\def\beqa{\begin{eqnarray}}
\def\eeqa#1{\label{#1}\end{eqnarray}}
\def\eeqan{\end{eqnarray}}

\def\Dslash{\not{\hbox{\kern-4pt $D$}}}
\def\dslash{\not{\hbox{\kern-2pt $\del$}}}

\def\msb{{\bar{\ssstyle M \kern -1pt S}}}

\usepackage{lineno}

\usepackage[intlimits,fleqn]{mathtools}
\usepackage{amssymb}
\usepackage{physics}
\usepackage{bm}

\usepackage[british]{babel}
\usepackage[babel=true]{microtype}
\usepackage[hidelinks]{hyperref}
\usepackage[noabbrev]{cleveref}
\usepackage[autostyle]{csquotes}
\usepackage{subcaption}
\usepackage{siunitx}
\newcommand\pubnumber{ PROC-CTD19-134 }

\newcommand\pubdate{30th May 2019}

\def\affiliation{
On behalf of the ATLAS Collaboration,\\
Physikalisches Institut\\
Albert-Ludwigs-Universität Freiburg, Germany}

\def\support{\footnote{Work supported by \enquote{Deutsche Forschungsgemeinschaft} and
\enquote{Bundesministerium für Bildung und Forschung}.\\
Copyright 2019 CERN for the benefit of the ATLAS Collaboration. CC-BY-4.0 license.}}

\begin{document}


\large
\begin{titlepage}
\pubblock

\vfill
\Title{Sensor shapes and weak modes of the ATLAS Inner Detector track-based alignment}
\vfill

\Author{Julian Wollrath \support}
\Address{\affiliation}
\vfill

\begin{Abstract}
	The alignment of the ATLAS Inner Detector is performed with a track-based algorithm. The aim of the
	detector alignment is to provide an accurate description of the detector geometry such that track
	parameters are accurately determined and bias free.
	The detector alignment is validated and improved by studying resonant decays ($J/\psi$ and $Z$ to
	$\mu^+\mu^-$). The detailed study of these resonances (together with the properties of the tracks of
	their decay products) allows to detect and correct for alignment weak modes such as detector curls
	and radial deformations that may bias the momentum and/or the impact parameter measurements. Here,
	radial distortions were investigated.
	Furthermore, a new analysis with a detailed scrutiny of the track-to-hit residuals allowed to study
	the deformation shape of the Pixel and IBL modules. The sensor distortion can result in track-to-hit
	residual biases of up to \SI{10}{\micro\metre} within a given module. The shape of the IBL modules
	was parametrised with Bernstein-Bézier functions and used to correct the hit position in the track
	fitting procedure.
\end{Abstract}

\vfill

\begin{Presented}
Connecting the Dots and Workshop on Intelligent Trackers (CTD/WIT 2019)\\
Instituto de F\'isica Corpuscular (IFIC), Valencia, Spain\\ 
April 2-5, 2019
\end{Presented}
\vfill
\end{titlepage}
\def\thefootnote{\fnsymbol{footnote}}
\setcounter{footnote}{0}
%

\normalsize 


\section{Introduction}\label{sec:intro}

Precision measurements such as that of the $W$ boson mass \cite{Aaboud:2017svj} at the ATLAS experiment
\cite{Aad:2008zzm} require a precise determination of the absolute momentum scale of charged particles
measured by the ATLAS Inner Detector (ID). Measurements of track momenta by the ID can be affected by
several sources of biases, one of them being residual geometric deformations of the ID prevailing after
its alignment, which can be either real geometrical deformations or ill-defined alignment solutions.
Here, new work on the assessment of radial distortions with a layer inflation model is described.

The alignment and the quality of the reconstructed tracks can also be influenced by imprecise knowledge
of the detector components. The extraction of the shape of ATLAS Insertable B-Layer (IBL) sensors
from track-to-hit residuals is also described here.

\section{Inner detector alignment}

The actual geometry of the detector differs from its nominal geometry and is determined by a track-based
alignment algorithm which is based on minimising track-to-hit residuals with the least square method
\cite{ATL-PHYS-PUB-2015-031}. The alignment parameters are determined iteratively and the alignment is
performed on different hierarchical levels, where either whole structures such as the different layers or
single modules are aligned. The alignment of different substructures or modules can introduce systematic
biases which can be assessed by reconstructing invariant masses of particles decaying into a pair of
muons, i.\,e. $J/\psi \rightarrow \mu^+\mu^-$ and $Z \rightarrow \mu^+\mu^-$.

\section{Radial distortions}\label{sec:radial_distortions}

The transverse momentum of a charged particle emerging from the centre of a cylindrical detector
satisfies the relation
\begin{linenomath}
\begin{equation}\label{eq:pT}
	p_\mathrm{T} \sim qB \frac{R_0^2}{8s},
\end{equation}
\end{linenomath}
with particle charge $q$, uniform magnetic field $B$, detector radius $R_0$ and sagitta $s$ of the
trajectory.

\begin{figure}
	\centering
	\includegraphics[width=.6\linewidth]{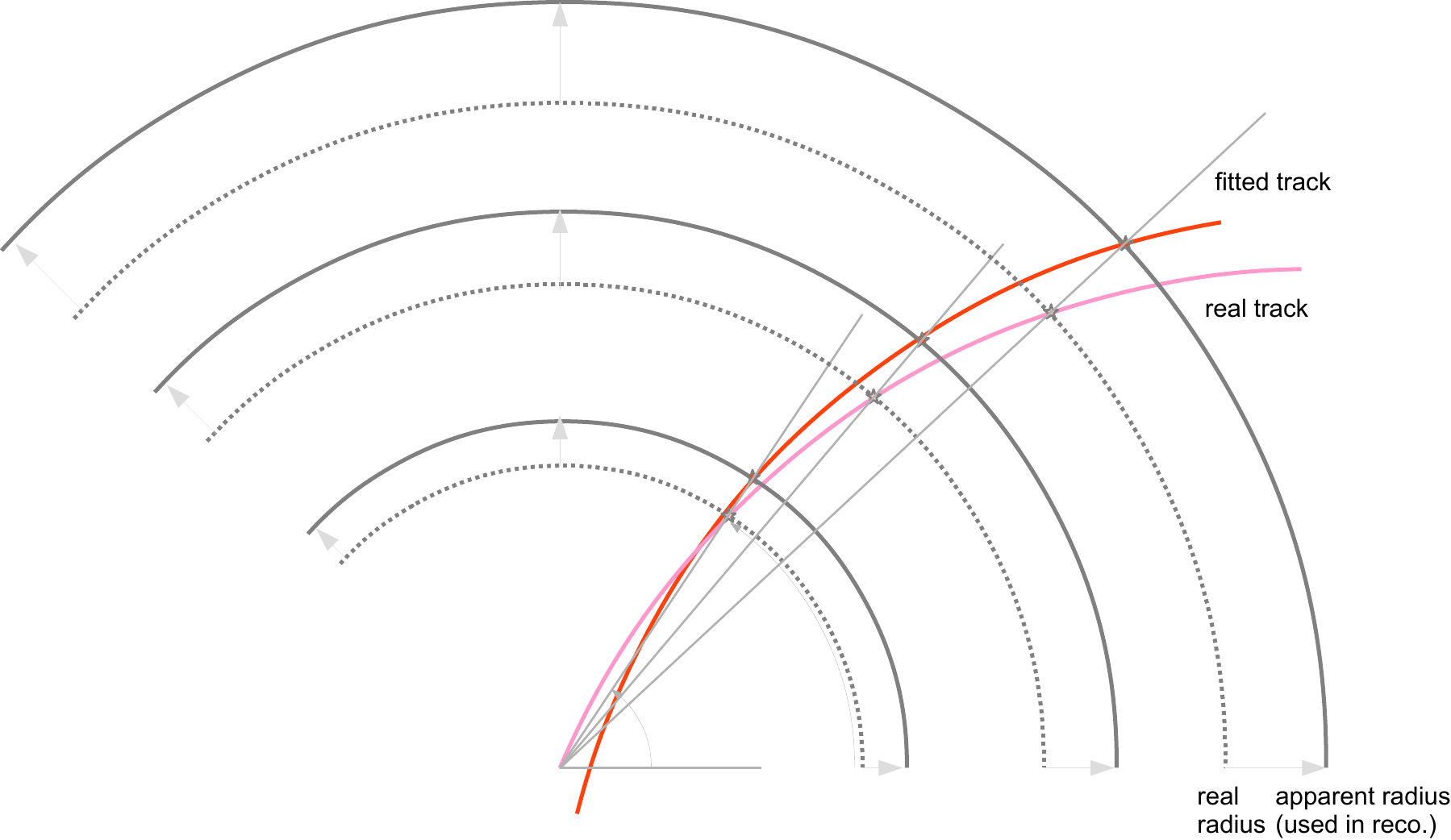}
	\caption{Sketch of a radial deformation: the real radii of the layers are shown as dotted lines,
	while the apparent layer radii are shown as solid lines. The inflation of the radius leads to a
	track-to-hit residual and a change in track curvature, which changes the momentum measurement.}
	\label{fig:layer_inflation}
\end{figure}

A geometrical deformation that alters the detector radius to $\tilde{R}_0 = R_0 + \delta R$ is called a
radial distortion and is depicted in \cref{fig:layer_inflation}. Radial distortions were previously
studied in ATL-PHYS-PUB-2018-003 \cite{ATL-PHYS-PUB-2018-003}, the present work focuses on their dependence on the azimuth
$\phi$ of the ID. It is assumed that the transverse and longitudinal momentum ($p_\mathrm{T}$ and
$p_\mathrm{z}$) of the tracks get changed to
\begin{linenomath}
\begin{equation}
	\tilde{p}_\mathrm{T} = p_\mathrm{T} \qty(1 + \frac{\delta R}{R_0})
\end{equation}
\end{linenomath}
and analogously for $p_\mathrm{z}$, while the incident angle $\theta$ of the tracks stay invariant. Given
that $p_{\mathrm{T},1/2}$ and $\eta_{1/2}$ denote transverse momentum and pseudorapidity of the
respective, assumed to be massless, decay particle, the mass $m$ of a particle undergoing a two-body decay
can be calculated as
\begin{linenomath}
\begin{equation}
	m^2 = 2 p_{\mathrm{T},1} p_{\mathrm{T},2} \qty\big(\cosh\qty(\eta_1 - \eta_2) - \cos\Delta\phi).
\end{equation}
\end{linenomath}
Hence, a radial distortion will change the measured mass to
\begin{linenomath}
\begin{equation}
	\tilde{m} = m \qty(1 + \frac{\delta R}{R_0}).
\end{equation}
\end{linenomath}



\begin{figure}
	\begin{subfigure}{.5\linewidth}
		\centering
		\includegraphics[width=\linewidth]{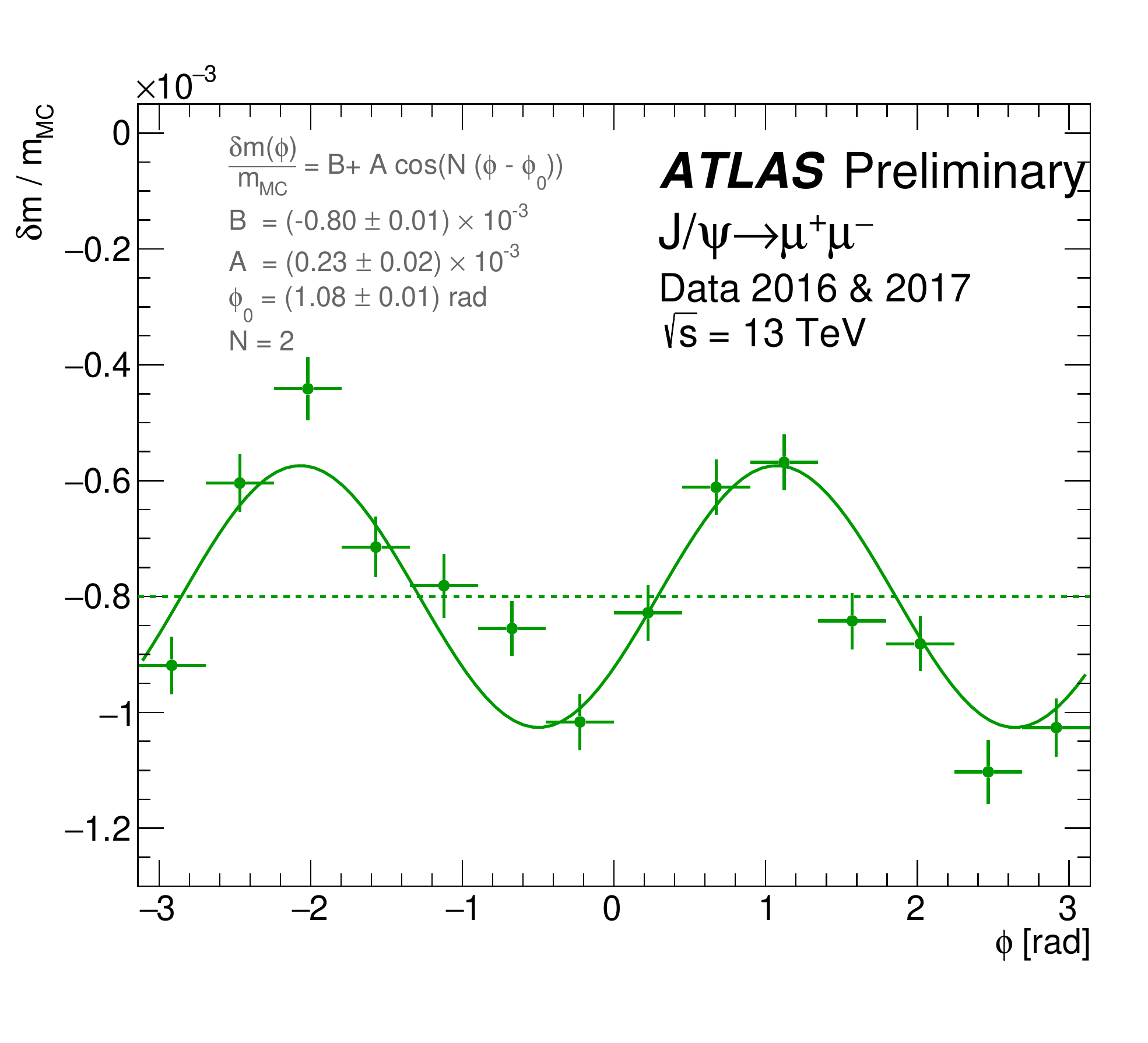}
	\end{subfigure}
	\begin{subfigure}{.5\linewidth}
		\centering
		\vspace{5pt}
		\includegraphics[width=\linewidth]{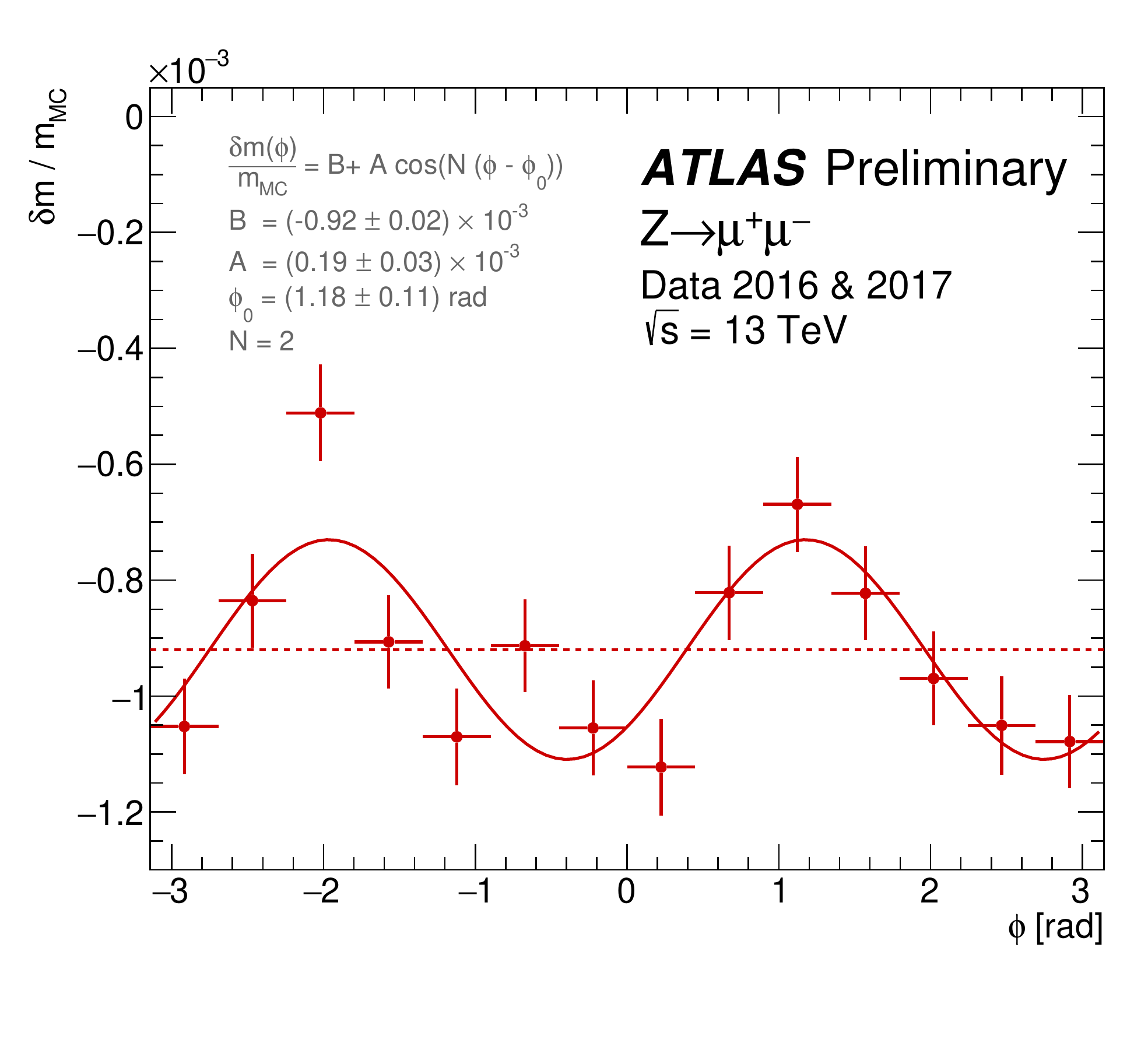}
	\end{subfigure}
	\caption{The reconstruction of invariant masses for $J/\psi \rightarrow \mu^+\mu^-$ (left) and $Z
	\rightarrow \mu^+\mu^-$ (right) points to the presence of a $\phi$-dependent radial distortion.}
	\label{fig:raddist_invmass}
\end{figure}

To assess the magnitude of this bias, the peak of the distribution of reconstructed di-muon masses
$m_\mathrm{fit}$ was compared to the peak of the distribution of the di-muon masses in Monte-Carlo
simulation $m_\mathrm{MC}$. The ratio
\begin{linenomath}
\begin{equation}
	\frac{\delta m}{m_\mathrm{MC}} = \frac{m_\mathrm{fit} - m_\mathrm{MC}}{m_\mathrm{MC}}
\end{equation}
\end{linenomath}
which relates to a radial distortion was plotted against $\phi$ as shown in \cref{fig:raddist_invmass}.
The data points at the presence of a $\phi$-dependent radial distortion, which could be observed in both
$J/\psi \rightarrow \mu^+\mu^-$ and $Z \rightarrow \mu^+\mu^-$.


\begin{figure}
	\centering
	\includegraphics[width=.6\linewidth]{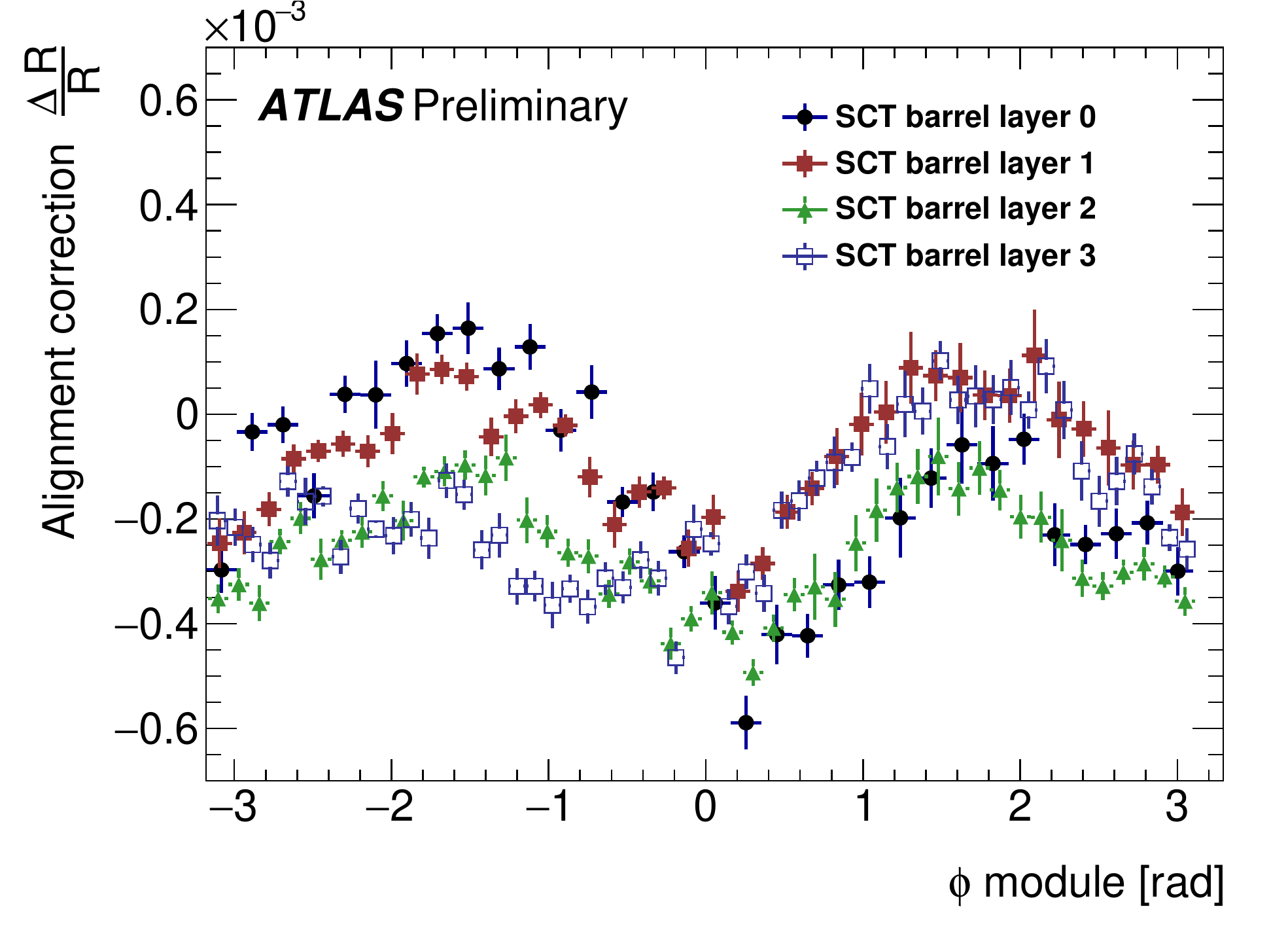}
	\caption{Radial distortions for the four SCT barrel layers. The modulation with respect to
	$\phi$ is compatible with an elliptical deformation with the minor axis in the horizontal plane and
	difference between major and minor axis of $\sim\SI{100}{\micro\metre}$.}
	\label{fig:raddist_SCTbarrel}
\end{figure}

Radial distortions were also evaluated for the four barrel layers of the silicon strip detector (SCT)
\cite{Aad:2014mta}. For these layers a radial distortion was also observed as shown in
\cref{fig:raddist_SCTbarrel}. These $\phi$-dependent radial distortions are compatible with an elliptical
deformation of the SCT layers with the minor axis of the deformation being in the horizontal plane and a
difference between minor and major axis of $\sim\SI{100}{\micro\metre}$.

\section{Insertable B-Layer sensor shape}\label{sec:IBL_shape}

\begin{figure}
	\begin{subfigure}{.5\linewidth}
		\centering
		\includegraphics[width=\linewidth]{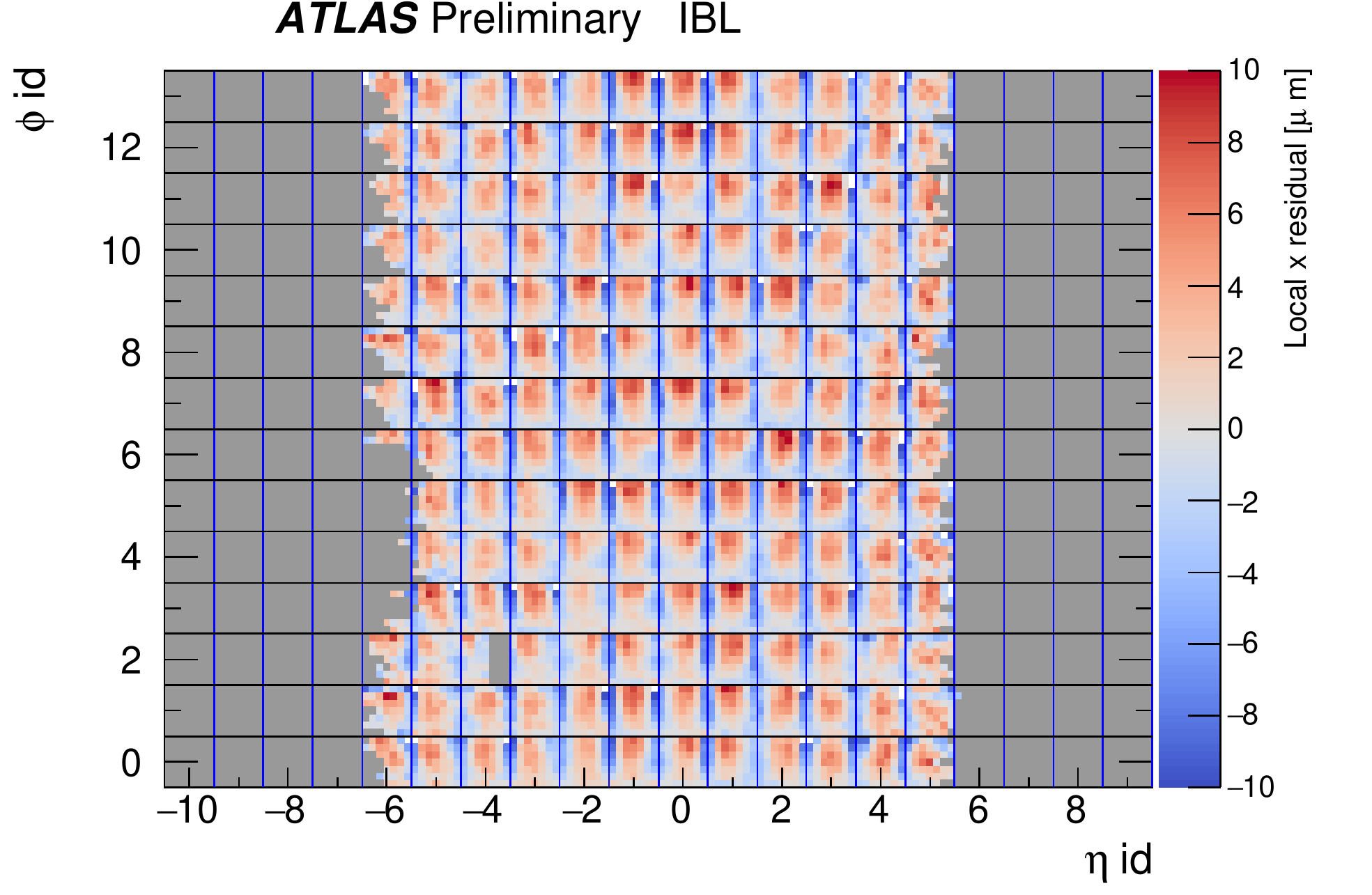}
	\end{subfigure}
	\begin{subfigure}{.5\linewidth}
		\centering
		\includegraphics[width=\linewidth]{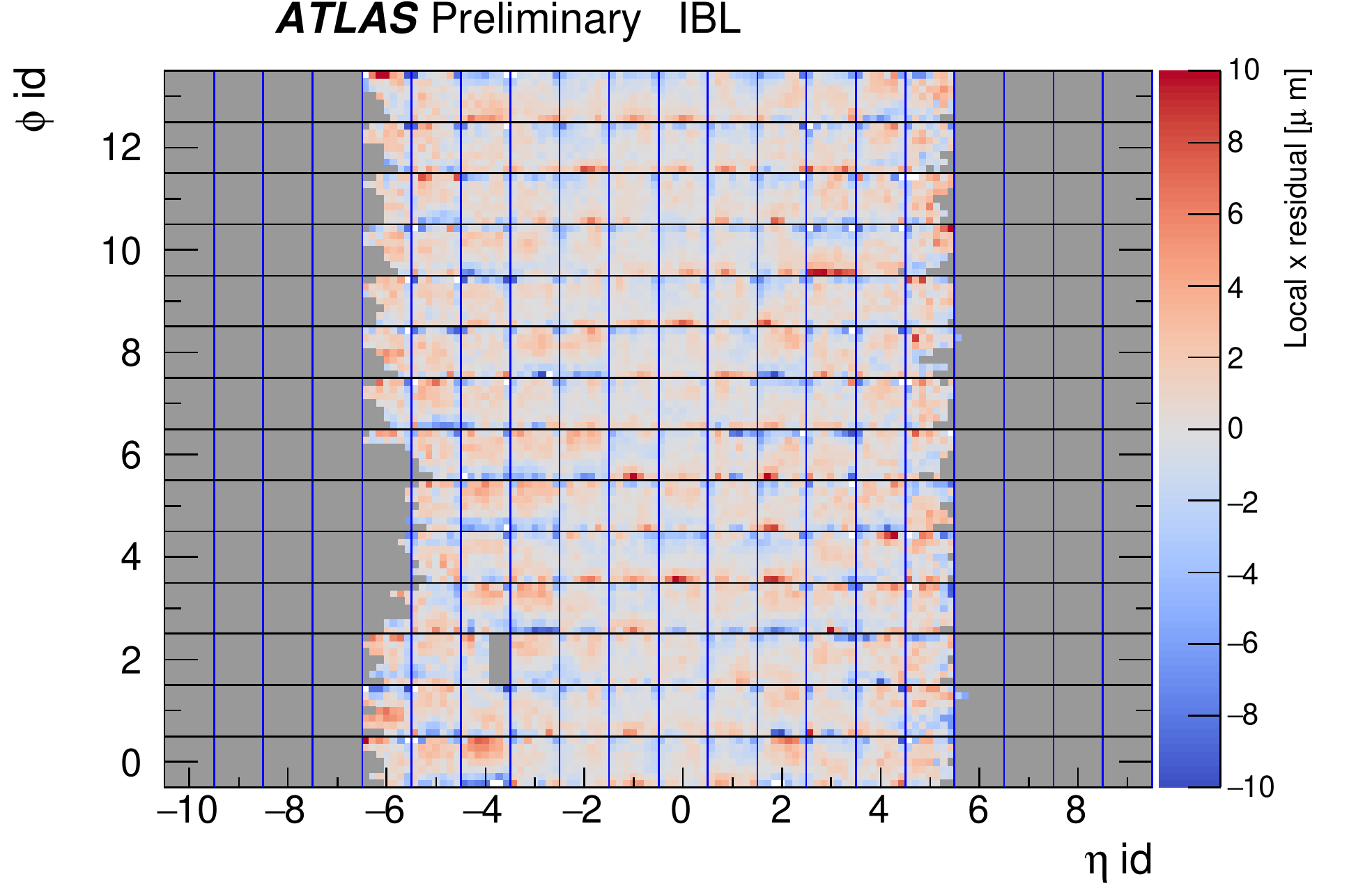}
	\end{subfigure}
	\begin{subfigure}{.5\linewidth}
		\centering
		\includegraphics[width=\linewidth]{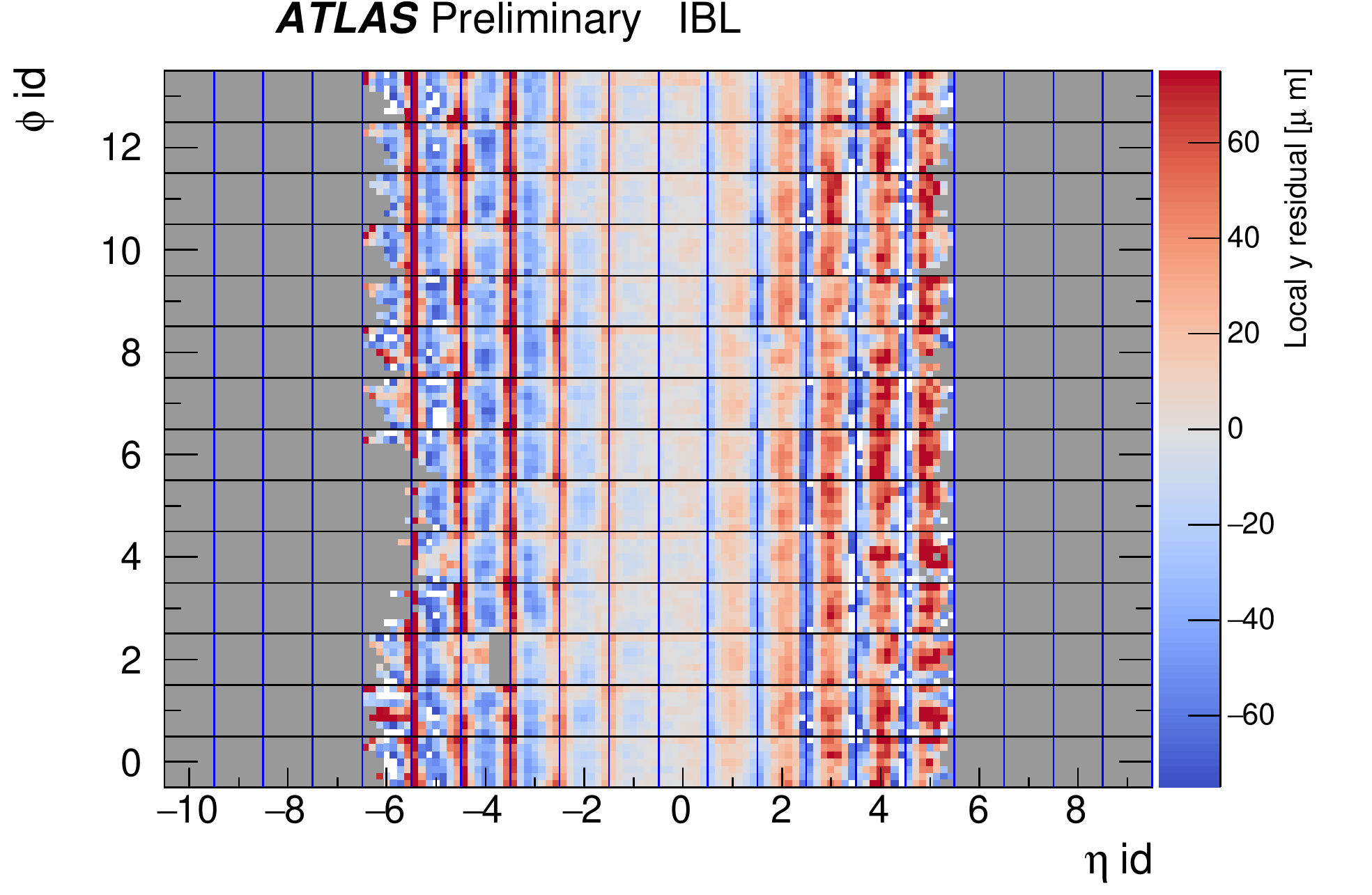}
	\end{subfigure}
	\begin{subfigure}{.5\linewidth}
		\centering
		\includegraphics[width=\linewidth]{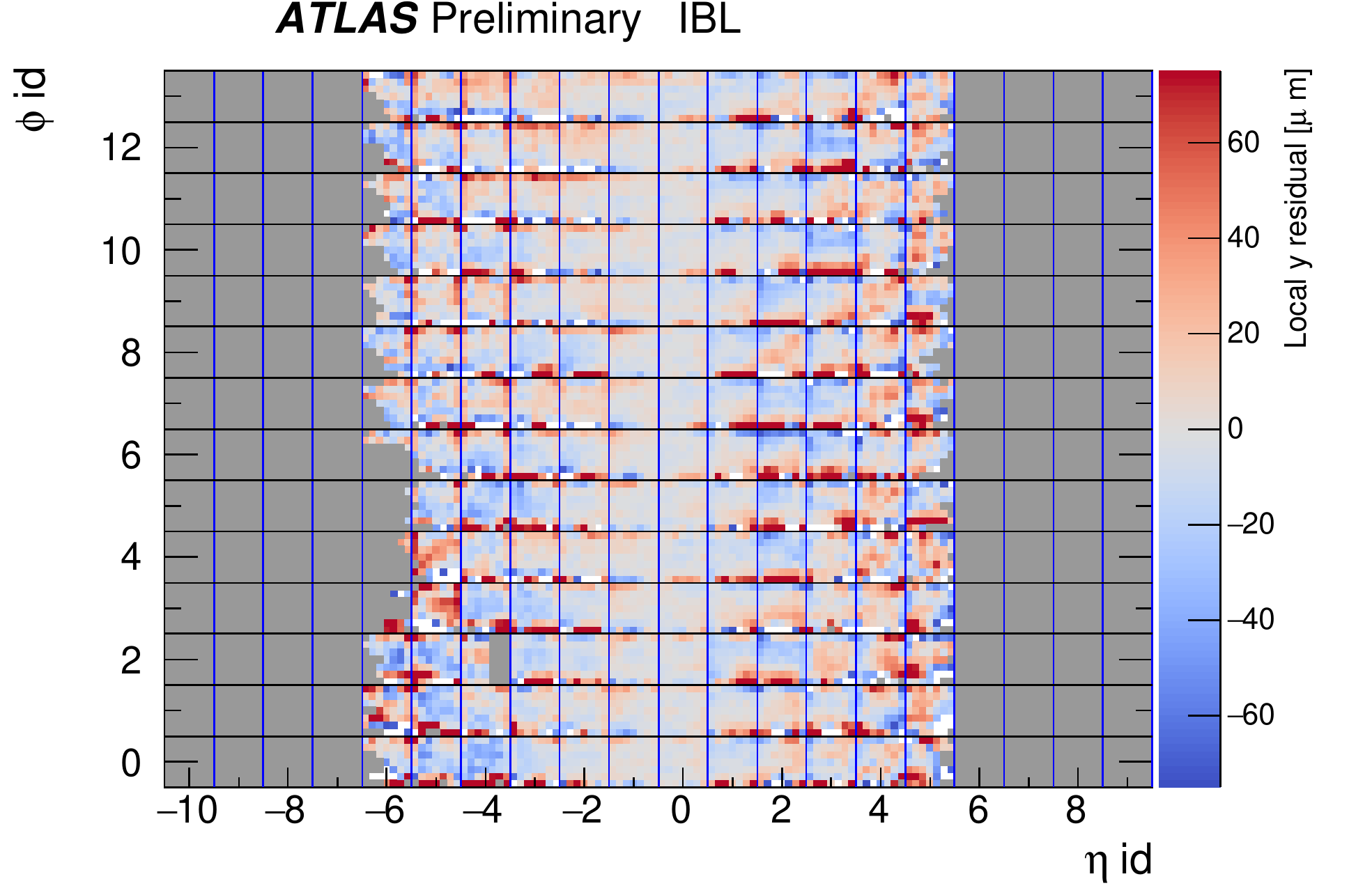}
	\end{subfigure}
	\caption{Intra-module local-$x$ (top) and local-$y$ (bottom) residuals for the IBL. On the left, IBL
	modules are assumed to be flat, while on the right shape corrections are taken into account. Taking
	the module shape into account reduces the intra-module residuals.}
	\label{fig:IBL_residuals}
\end{figure}

The IBL \cite{Capeans:2010jnh}, which was installed in 2014, consists of 14 staves. From one end to the
other each of these is equipped with four 3D modules, twelve planar modules and again four 3D modules;
all of these modules are silicon pixel modules. While the shapes of the modules of the original pixel
detector were measured in a campaign in 2006 before they were installed into the experiment and hence
their shape is taken into account when reconstructing tracks, no such measurement was done for the IBL
modules; therefore they are assumed to be flat in track reconstruction. Whether this assumption holds can
be evaluated by splitting the modules into small cells and calculating the average track-to-hit residual
in these cells. These intra-module track-to-hit residuals of the IBL planar modules were evaluated for
data from one long LHC fill taken on 9th to 10th November 2017 containing $\sim\num{9e5}$ IBL hits. This
was done by rerunning the track reconstruction and performing one iteration of alignment for the IBL and
pixel modules as is shown on the left hand side of \cref{fig:IBL_residuals}. While the average
track-to-hit residual per module is $\sim\SI{0}{\micro\metre}$, a clear intra-module structure, which is
similar for all the modules, could be observed. This means that the IBL modules cannot assumed to be flat
but instead have a shape in local-$z$; this shape can be extracted from the track-to-hit residuals and
the track parameters. For this shape extraction a second, larger statistics data sample from late 2017
containing $\sim\num{2e9}$ tracks was used. Since the resolution of the detectors is finer in local-$x$
than in local-$y$, the $z$-value for each track with an IBL hit was calculated as
\begin{linenomath}
\begin{equation}
	z' = x_\mathrm{residual} \cot\phi'.
\end{equation}
\end{linenomath}
The sensors were than split into $21 \times 21$ cells and the local-$z$ value of each cell
$\mathfrak{P}_{i,j}$ was calculated as the weighted mean of the $z'$ values in the cell. To get a smooth
approximation of the sensor surface this extracted shape was subsequently interpolated with a
two-dimensional Bernstein-Bézier function as defined in \cref{eq:bernsteinbeziersurface} with the
Bernstein basis polynomials defined in \cref{eq:bernsteinbezierbasis}.
\begin{linenomath}
\begin{gather}\label{eq:bernsteinbeziersurface}
	\mathfrak{r}\qty(u, v) = \sum_{i = 0}^{n} \sum_{j = 0}^{m} B_{i, n}\qty(u) B_{j,m}\qty(v)
	\mathfrak{P}_{i,j}\ ,\quad 0 \leq u,v \leq 1\\
    \label{eq:bernsteinbezierbasis}
	B_{i,n} = \binom{n}{i} t^i \qty(1 - t)^{n - i}\ ,\quad
	0 \leq t \leq 1\quad\qty(i = 0, 1,\dotsc, n)
\end{gather}
\end{linenomath}

\begin{figure}
	\begin{subfigure}{.5\linewidth}
		\centering
		\includegraphics[width=\linewidth]{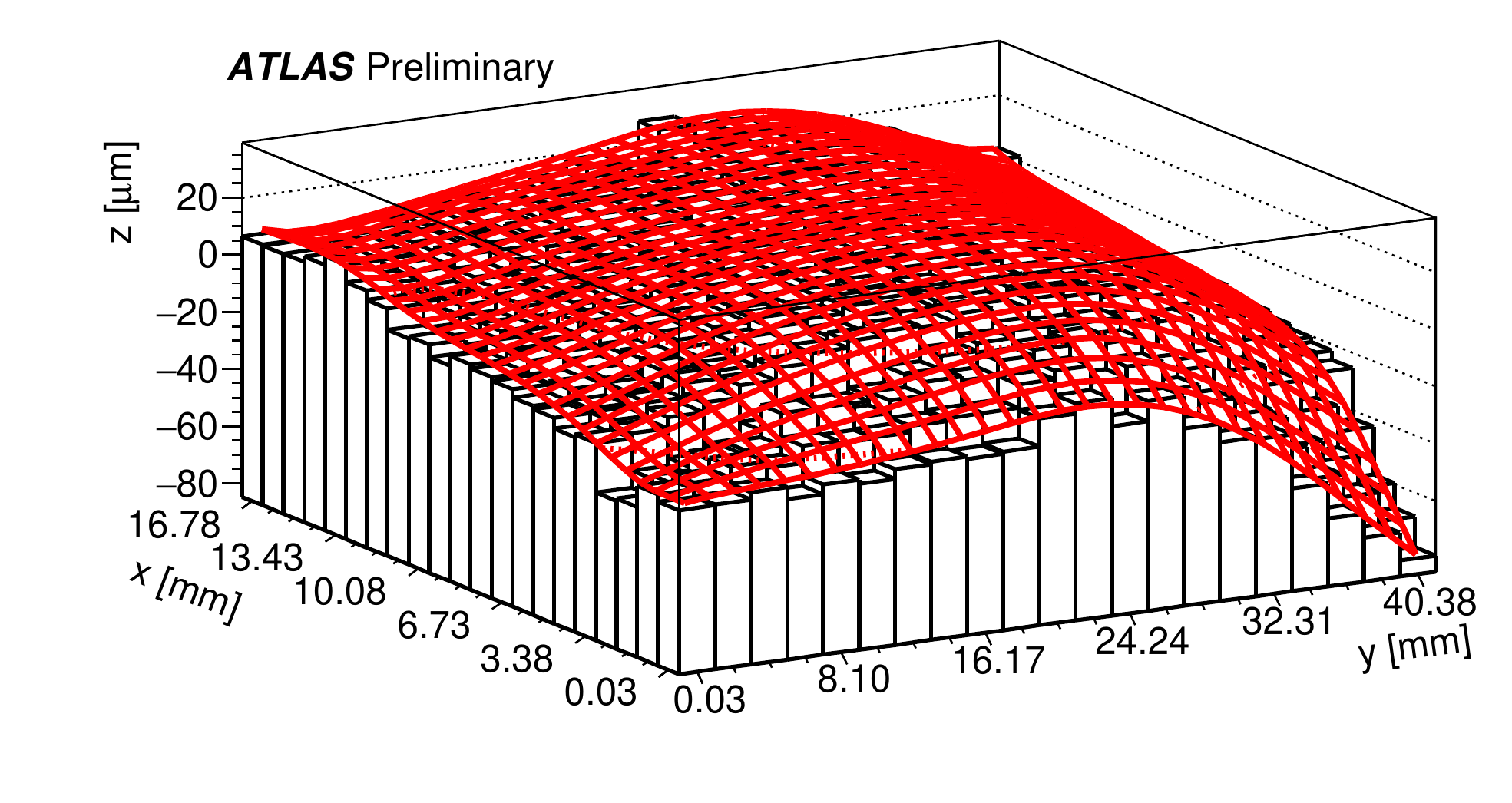}
	\end{subfigure}
	\begin{subfigure}{.5\linewidth}
		\centering
		\includegraphics[width=\linewidth]{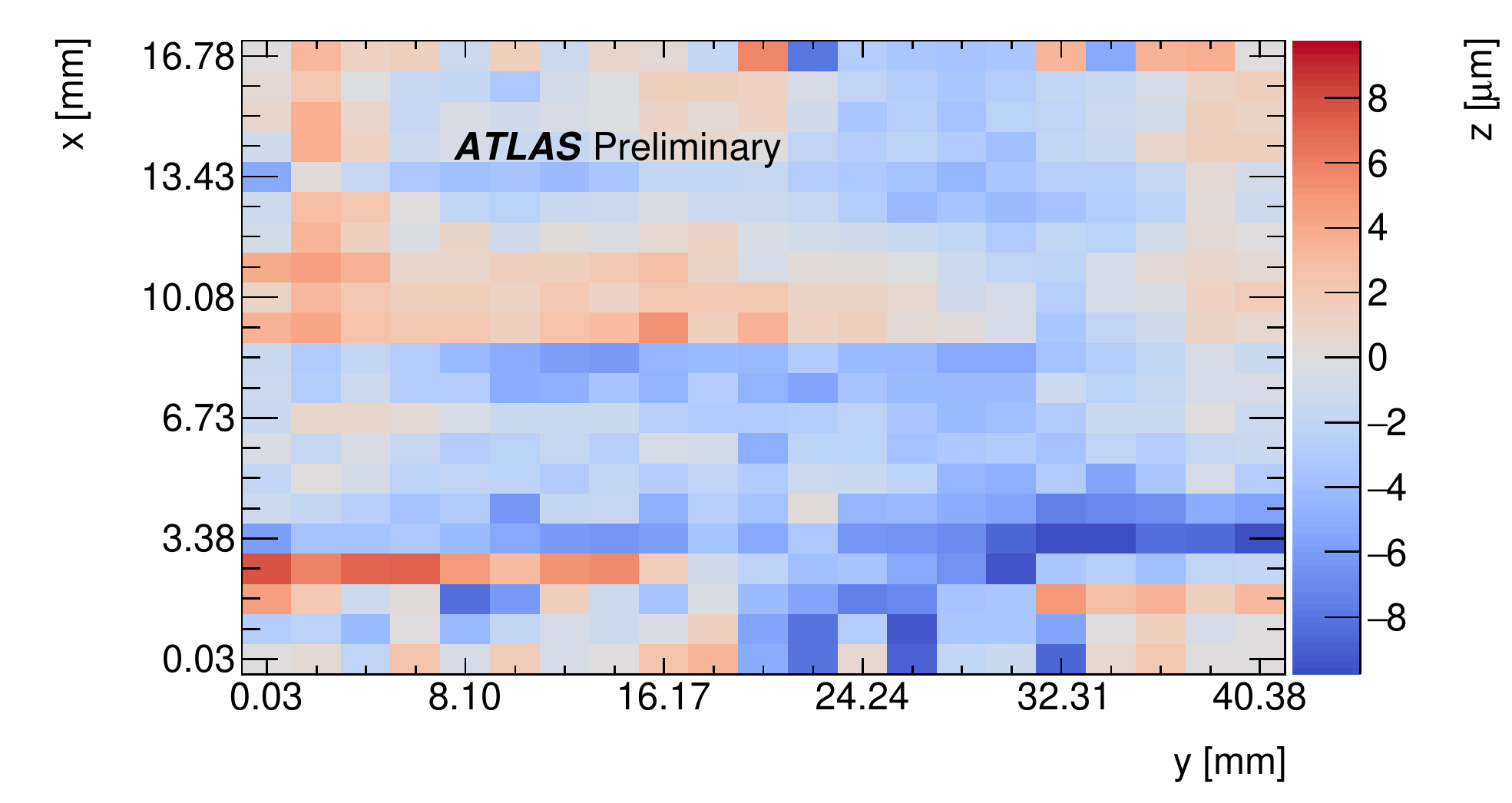}
	\end{subfigure}
	\caption{The shape of the IBL module with $\phi$-ID 10 and $\eta$-ID 0 was extracted from
	track-to-hit residuals (left, black) and interpolated with a Bernstein-Bézier function (left, red).
	The difference between the shape extraction (black) and the interpolation (red) is shown on the
	right.}
	\label{fig:IBL_module_shape}
\end{figure}

Utilising the data sample with a huge number of tracks, the shape of the planar IBL modules was measured.
This shape and its interpolation for the IBL module with $\phi$-ID 10 and $\eta$-ID 0 is shown on the
left hand side of \cref{fig:IBL_module_shape} and the difference between measured shape and interpolation
is shown on the right hand side of \cref{fig:IBL_module_shape}; for most of the sensor, the deviation
between measurement and interpolation is in the $\pm\SI{2}{\micro\metre}$ range.

Over most of the modules, the sensor bends in the $\pm\SI{20}{\micro\metre}$ range. A clear trend towards
large negative values of the bending is observed for smaller $x$, which might be due to stress of a
kapton flex that is attached to the module on that side.

To evaluate the impact of including the shape information for the IBL in the reconstruction, one iteration
of alignment of the IBL and pixel modules was performed for the same LHC fill as before but this time taking
the sensor shape into account. The intra-module track-to-hit residuals were evaluated as shown on the
right hand side of \cref{fig:IBL_residuals}. A clear improvement could be observed in the intra-module
local-$x$ and local-$y$ residuals with the structure in the local-$x$ residuals being mostly removed.

\section{Conclusions}\label{sec:conclusions}

Radial distortions of the ATLAS Inner Detector were studied with a layer inflation model. The
reconstruction of invariant masses point to the presence of a $\phi$-dependent radial distortion.
Furthermore the SCT barrel layers show a $\phi$-dependent radial distortion compatible with an elliptical
deformation. Having assessed these distortions, the knowledge about them could be included into the
alignment procedure as an additional constraint and help in improving the ID alignment.

The shape of IBL sensors was extracted from track-to-hit residuals. A clear deviation from the flat
sensor hypothesis, mostly in the $\pm\SI{20}{\micro\metre}$ range, was observed. Taking the shape of the
sensors into account removed most of the structure observed in the local-$x$ and local-$y$ residuals.
Adding the correct shape into the track reconstruction will possibly help in improving the precise
determination of track parameters such as the impact parameter.





\end{document}